\documentclass[a4paper,11pt]{article}%
\usepackage{graphicx}
\graphicspath{{Figures/}} 
\usepackage{amsmath}
\usepackage{amssymb}
\usepackage{subcaption}
\usepackage[colorlinks]{hyperref}
\usepackage{color}%
\usepackage{cite}
\usepackage{float}

\newcommand{\bra}{\begin{array}}
\newcommand{\era}{\end{array}}
\newcommand{\beq}{\begin{equation}}
\newcommand{\eeq}{\end{equation}}
\newcommand{\bqr}{\begin{eqnarray}}
\newcommand{\eqr}{\end{eqnarray}}

\def\BC{\bb C}
\def\_\BC{\bbi C}

\def\( {\left(}
   \def\) {\right)}
\def\[ {\left[}
\def\] {\right]}
\def\no2 {{\textstyle{n\over 2}}}

\def\dag {{\dagger}}

\newcommand{\si}{\sigma}

\newcommand{\pa}{\partial}

\newcommand{\app}{\approx}
\newcommand{\del}{\delta}
\newcommand{\lga}{\longrightarrow}

\newcommand{\da}{\dagger}

\newcommand{\lb}{\label}

\begin{document}
\begin{titlepage}
\setcounter{page}{1}
\renewcommand{\thefootnote}{\fnsymbol{footnote}}

\begin{flushright}
\end{flushright}

\vspace{5mm}
\begin{center}

{\Large \bf
Measuring Space Deformation via Graphene under Constraints
}

\vspace{5mm}

{\bf Ahmed Jellal}\footnote{\sf{a.jellal@ucd.ac.ma}}

{
\em Laboratory of Theoretical Physics,  
Faculty of Sciences, Choua\"ib Doukkali University},\\
{\em PO Box 20, 24000 El Jadida, Morocco}

{
\em Canadian Quantum  Research Center,  
204-3002 32 Ave Vernon, \\ BC V1T 2L7,  Canada}

\vspace{3cm}

\begin{abstract}

We describe the lattice deformation in graphene under strain effect by considering 
the spacial-momenta  coordinates do not commute. 
This later can be realized by 
introducing the star product to end up with a generalized Heisenberg algebra.
Within such framework, we build a new model describing Dirac fermions 
interacting with  an external source that noncommutative parameter $\kappa$ dependent. The solutions of the energy spectrum are showing  effective Landau levels  in similar way to the case of a real magnetic field applied to graphene. We show that some 
strain configurations would be able to explicitly evaluate $\kappa$
and then  offer a piste toward  its measurement.

\vspace{5cm}

\noindent PACS numbers:  71.15.-m, 02.40.Gh

\noindent Keywords: Graphene, noncommutative coordinates, Dirac equation, strain effect.

\end{abstract}

\end{center}
\end{titlepage}

\section{Introduction}

Graphene  was experimentally isolated in 2004 and is 
a sheet of carbon atoms arranged in
hexagonal cells only a single atom thick \cite{Novoselov}. It presents exostic
properties in particular
a linear dispersion relation, which is quite different to that of semi-conductors.
This relation
can be quantized under the influence of 
magnetic field applied to graphene sheet 
giving rise to the Landau levels \cite{Castro}. Such 
field can be real or 
a manifestation of the graphene honeycomb under certain strain patterns generating the so-called pseudomagnetic field \cite{Vozmediano,Guinea, Wijk}. In fact, it was shown that
 the
strained-induced pseudomagnetic field can easily reach
quantizing values, exceeding $10$ T in submicron devices for
deformations less than $10 \%$ \cite{Guinea2}. Besides
the pseudomagnetic field can be generated in graphene
through wafer-scale epitaxial growth or  shallow triangular nanoprisms in the SiC substrate \cite{Nigge}.
%
The importance of 
the quantized magnetic fields 
resides in the emergence of the anomalous quantum Hall effect \cite{Novoselov2005, Zhang2005}. Also it establishes  a link with important theorems of modern mathematical physics 
as an example the existence of the zero-energy Landau level, which is a consequence of  
the Atiyah-Singer index theorem \cite{Atiyah, Katsnelsonbook12}.


On the other hand, 
the deformation based on the noncommutative geometry \cite{Connes} is  a power mathematical tool
to deal with different issues in physics. Many works 
were reported on quantum mechanics in hopping to find some extensions and corrections to the known standard results. To this end, we may refer to our previous works 
on the quantum Hall effect \cite{Jellal1} or statistical mechanics \cite{Jellal2} as well other issues \cite{Jellal3}.
One way to invent the noncommutativity in a given physical system is to make use start product giving rise to a generalized Heisenberg algebra that involves commutation relations between positions and momenta coordinates
in addition to the standard relations. Strictly speaking,
 noncommutativity can be imposed by treating the coordinates
as commuting but requiring that composition of their functions is given in terms of the star product. In particular, it is shown that  the noncommutative features affected the Landau levels
by offering some corrections to the solutions of the energy spectrum
either only taking the spatial-spatial being noncommute or 
both spatial-momenta do not commute, one may see \cite{Jellal1, Jellal2,Jellal3} and references therein.

With respect to the extended Landau levels on the noncommutative plane, we propose
a newly method to create a bridge between the noncommutativity of the coordinates and deformed graphene through the strain effect.
To achieve this goal, we 
consider 
the Dirac fermions in pristine graphene moving in plane by taking 
spatial-momenta do not commute. The resulting Hamiltonian shows some similarities with that of the Landau problem for one particle experiencing a perpendicular magnetic field. 
This allows us 
to easily derive a quantized energy spectrum showing a dependency on the noncommutative parameter $\kappa$. To interpret our results, we offer two possibilities either by linking $\kappa$
to an effective magnetic field or expressing $\kappa$ in terms of the strain tensor components.
To this end, 
we provide a way to measure $\kappa$ 
and  
fix its value by evaluating different physical parameters. 

The present paper is organized as follows. In section 2, we build a theoretical model that captures the noncommutativity effect and leads to a Hamiltonian involving an effective vector potential taken  
in the symmetric gauge. The diagonalization of such Hamiltonian gives rise to
%
effective Landau levels and  eigenspinors similar to those of fermions in graphene subject to  
a perpendicular magnetic field. 
Naturally,  these levels disappear and reduce to a linear dispersion relation by switching off $\kappa$. 
In section 3, we propose a method to shed light on the measurement of $\kappa$ by considering the strain tensor patterns in graphene and then offer  numerical  values. We conclude our results in the final section.

\section{Building  model}

Graphene can be described by a low-energy
continuum model involving a matrix $4\times 4$
because it 
possess two-Fermi points and each one has 
a two-fold band degeneracy. 
%
Such matrix has
 four-component envelope
wavefunction whose components are labeled by a Fermi-point
pseudospin $=\pm1$ and a sublattice forming an honeycomb.
Then  one-pseudospin Hamiltonian for a free Dirac fermion 
reads as
%
\beq\lb{ham1}
H= v_{\sf F} {\vec\sigma}\cdot \vec p
\eeq
where $\vec \si=(\si_x,\si_y)$ are the Pauli matrices, 
$\vec p=(p_x, p_y)$ is the momenta and  $v_{\sf F}\app 10^6$ m/s is the Fermi velocity. One can easily show that
the solution of the energy spectrum has  
a linear dispersion relation  $E=\pm v_{\sf F} k$ with 
 the wave vector $\vec k=(k_x,k_y)$. 
 Now by applying a perpendicular magnetic $B$ to graphene sheet and choosing
 the symmetric gauge 
 $\vec A_{B}=\frac{B}{2}\left(y, -x\right)$, the  Hamiltonian \eqref{ham1} 
 can be transformed to 
 \beq\lb{ham22}
H_{B}= v_{\sf F} \left[\sigma_x \left(p_x + \frac{Be}{2} y\right)
+ \sigma_y \left(p_y - \frac{Be}{2} x\right)\right]
\eeq
 having quantized energy $E_n= \pm v_{\sf F} \sqrt{2Bn}$ with $n$ is an integer value. In the next, we will apply a mathematical machinery based on the noncommutativity of spacial-momenta coordinates  
 to obtain similar results to \eqref{ham22}. 
 

To deal with our task, we consider  Dirac fermions living on 
 the noncommutative space
and impose  the noncommutativity  by treating the phase space coordinates
as commuting but 
composing 
their functions with respect to the star product. Then, this process yields to
the deformed Heisenberg algebra described by the commutation relations
\bqr
 [x_j, p_k]= i\hbar \del_{jk}, \qquad  [x_j, x_k]= i\theta \del_{jk}, \qquad
 [p_j, p_k]= i\kappa \del_{jk}
\eqr
where $\theta$ and $\kappa$
are two antisymmetric noncommutative parameters measuring the noncommutativity between position $x_i$ and momenta $p_i$
coordinates. They play the same role as for the Planck constant 
in the canonical commutation relation between $x_i$ and $p_i$.
Now taking into account of the noncommutativity, we can define
an another Hamiltonian using the action of the Hamiltonian \eqref{ham1} via the start product
on the spinor $\psi(x,y)$, namely
\beq
H\star\psi(x,y)=H_{\kappa} \psi(x,y)
\eeq
which leads to  the following mappings
\beq\lb{pxy}
p_x\lga p_x + \frac{\kappa}{2\hbar} y, \qquad
p_y\lga p_y - \frac{\kappa}{2\hbar} x
\eeq
 and consequently we end up with 
the noncommutative Dirac Hamiltonian 
\beq\lb{ham2}
H_{\kappa}= v_{\sf F} \left[\sigma_x \left(p_x + \frac{\kappa}{2\hbar} y\right)
+ \sigma_y \left(p_y - \frac{\kappa}{2\hbar} x\right)\right]
\eeq
At this level, 
we have some comments in order. Indeed, 
\eqref{ham2}
looks like the Hamiltonian \eqref{ham22} for a real magnetic field associated
to the symmetric gauge.
%
This 
  suggest to define an effective 
gauge  $\vec A_{\kappa}=\frac{\kappa}{2\hbar}\left(y, -x\right)$ and then write $H_{\kappa}$  in the compact form
\beq
H_{\kappa}= v_{\sf F} \vec \si\cdot (\vec p+ e\vec A_\kappa)
\eeq
and therefore by analogy to the  relation $\vec B=\nabla \times \vec A$, we can define
an effective magnetic field linked 
to the noncommutative parameter $\kappa$ via the relation
%
\beq\lb{bkk}
B_\kappa= \frac{\pa A_{\kappa y}}{\pa x} - \frac{\pa A_{\kappa x}}{\pa y}= \frac{\kappa}{\hbar}
\eeq
which  has different consequences. For  instance 
it can be used to propose a new way to realize
the 
quantum Hall effect without need of a real magnetic field. In addition,
\eqref{bkk} allows to 
determine the solutions of the energy spectrum in similar way to that of the Landau problem in two-dimensions.


The Hamiltonian $H_{\kappa}$ can be diagonalized by 
 considering
the complex notation such that $z=x+iy$ and 
$2p_{z,\bar z} = p_x\mp i p_y$. Then, after some algebra, we obtain 
\beq
H_{\kappa}= v_{\sf F}
\left(
\begin{matrix}
 0 & 2p_z +i\frac{\kappa}{2\hbar} \bar z\\
 2p_{\bar z} -i\frac{\kappa}{2\hbar} z &0
\end{matrix}
\right).
\eeq
Such form suggests to
  linearly recombine the
complex variables in order to realize the bosonic ladder operators of the harmonic
oscillator. They are given by
%
\beq
a =\frac{1}{\sqrt{2\kappa}} \left(2p_{\bar z} -i\frac{\kappa}{2\hbar}z\right), \qquad
a^\da =\frac{1}{\sqrt{2\kappa}} \left(2p_{z} +i\frac{\kappa}{2\hbar}\bar z\right)
\eeq
which
satisfy the commutation relation $[a, a^\da]=\mathbb{I}$. Then,
the Hamiltonian can easily be mapped in terms of such operators as
\begin{equation}\label{hadm}
H_{\kappa}=v_{\sf F}\sqrt{2\kappa}\left(%
\begin{array}{cc}
  0 & \hat{a}^{\dag} \\
  \hat{a} & 0 \\
\end{array}%
\right).
\end{equation}
Therefore,
the solutions of the energy spectrum can be obtained
using 
the number operator $N = a^\da a$ together with the harmonic oscillator states. 
Indeed,
we solve the eigenvalue equation $H_{\kappa}\psi=E \psi$ to  end up with 
the quantized energy
in terms of the noncommutative parameter $\kappa$
\beq\lb{ekn}
E_{\kappa,n}= \pm v_{\sf F} \sqrt{2\kappa n}, \qquad n\in \mathbb{N}
\eeq
exhibiting 
 effective  Landau levels in similar way to  that of the Hamiltonian \eqref{ham22}. 
 This is showing
an effective zero-energy
Landau level, then one may talk here also about the consequence of the Atiyah-Singer index theorem as has been done in \cite{Katsnelsonbook12}.
The corresponding eigenspinors of \eqref{ekn} can be worked out using the  condition
 $a\psi_0=0$ to obtain the degenerate groundstate 
\beq\lb{fff}
\psi_0(z,\bar z)= \sqrt{\frac{\kappa}{2\pi\hbar}}  e^{-\frac{\kappa}{4\hbar} z\bar z}
\eeq
 and to get the full state we need to 
define another set of bosonic operators commuting with the former ones. They can be written as
\beq
b^\da =\frac{1}{\sqrt{2\kappa}} \left(2p_{\bar z} +i\frac{\kappa}{2\hbar}z\right), \qquad
b =\frac{1}{\sqrt{2\kappa}} \left(2p_{z} -i\frac{\kappa}{2\hbar}\bar z\right)
\eeq
and verify $[b, b^\da]= \mathbb{I}$. Calculating the action of the operator  $(b^\da)^m$
on \eqref{fff}, we find the normalized wavefunction
\beq
\psi_{0,m}(z,\bar z)= \sqrt{\frac{\hbar }{\pi \kappa 2^{m+1}
m!}}\left(\frac{\kappa }{\hbar} z\right)^{m}
e^{-\frac{\kappa}{4\hbar} z\bar z},\qquad
    m\in \mathbb{Z}
\eeq
and therefore the action $(a^\da)^n\psi_{0,m}(z,\bar z)$ gives
the normalized eigenspinors of the Hamiltonian $H_{\kappa}$ 
\begin{equation}\label{eigfuno}
    \Psi_{n\neq0,m}=\left(%
\begin{array}{c}
  \psi_{n, m} \\
\psi_{n-1, m} \\
\end{array}%
\right)
\end{equation}
where the eigenfunctions $\psi_{n,m}$
take the form
\begin{equation}\label{eigfun}
    \psi_{n,m} (z,\bar z)=\sqrt{\frac{\hbar n!}{\pi \kappa 2^{m+1}
(n+m)!}}\left(\frac{\kappa }{\hbar} z\right)^{m}
L^{m}_{n}\left(\frac{\kappa }{2\hbar} z\bar{z}\right )\ e^{-\frac{\kappa}{4\hbar} z\bar z}.
\end{equation}
Note that, the zero-mode eigenspinors is
\begin{equation}\label{zmod}
    \Psi_{0,m}=\left(%
\begin{array}{c}
  \psi_{0, m} \\
0 \\
\end{array}%
\right).
\end{equation}
To close this part, we notice that by switching off $\kappa$ we recover the linear dispersion relation and the associated eigenspinor. Additionally, the solutions of the energy spectrum for a real magnetic field can be obtained using the identification settled in \eqref{bkk}. 


\section{Measurement of $\kappa$}



Graphene has an hexagonal structure with  is  one-atom-thick film, which 
can be  considered as a flexible membrane. With
such characteristics, the connection between 
%
its mechanical and electronic properties is possible. Indeed, 
the interplay between its elastic and electronic
properties was investigated in different occasions \cite{Vozmediano, Guinea, Wijk}.
Experimental results showed that 
the graphene nanobubble under strain effect leads to a huge pseudomagnetic field (< 300 T),
which has never been created in the laboratory \cite{Levy}. Then, a uniaxial
strain larger than $23\%$ in the zigzag direction can generate a transport
gap in the transmission \cite{Fogler, Pereira}. Additionally, the mechanical strains in
graphene can also shift the Dirac points and generate new ones, 
which push  Dirac fermions to have asymmetrical velocity $v_x \neq v_y$
along $x$- and $y$-directions \cite{Choi}.
Motivated by the above results, we propose a way to measure the noncommutative parameter $\kappa$ by considering
 graphene under constraints. 

To clarify the above statement, we introduce 
the strain tensor for a given  
a spatially varying displacement field $\vec u= u_x\vec e_x+  u_y\vec e_y + h  \vec e_z$ in cartisian coordinates with
$u_i$ and $h$ are  the in and out of plane deformations.
Such tensor is given by
\beq
\varepsilon_{ij} = \frac{1}{2} \left(\pa_j u_i + \pa_i u_j + \pa_i h \pa_j h\right), \qquad i,j=x,y
\eeq
and then the structural distortion-induced strain on graphene
can be evaluated using 
 the following components
\begin{eqnarray}
&& u_{xx}= \frac{\pa u_x}{\pa x} + \frac{1}{2} \left(\frac{\pa h}{\pa x}\right)^2,
\qquad u_{yy}= \frac{\pa u_y}{\pa y} + \frac{1}{2} \left(\frac{\pa h}{\pa y}\right)^2\\
&&
u_{xy}= \frac{1}{2} \left(\frac{\pa u_x}{\pa x} + \frac{\pa u_y}{\pa y}\right)
+ \frac{1}{2} \left(\frac{\pa h}{\pa x} + \frac{\pa h}{\pa y}\right).
\end{eqnarray}
To explicitly determine these components, we work out with some
%
%
requirements related to graphene under strain effect. Indeed, 
 let us choose the $x$-direction such that
%
the structure deformation leads to 
 an effective gauge $\vec A_{\kappa}=\frac{\kappa}{2\hbar}(x,y) $  associated with the shear deformation at one site
 (in the first order approximation)
 \cite{Suzuura,  Manes}
 \begin{eqnarray}
 &&
  \frac{\kappa}{2\hbar}y=\pm
  \frac{c\beta}{a} (u_{xx} - u_{yy}) \lb{axx}\\
&& 
- \frac{\kappa}{2\hbar} x
= \mp 2\frac{c\beta }{a} u_{xy}  \lb{axy}
 \end{eqnarray}
where 
$\beta= -\pa \log t/\pa \log a$
is the electron Gr\"uneisen parameter, 
$a$ is the lattice spacing, 
 $t$ is the electron
hopping between $p_z$ orbitals located at nearest neighbor atoms,
$c$ is a numerical constant that depends on the details of atomic 
displacements within the lattice unit cell,  
the  signs $\pm$ refer to 
the $K$ and $K′$ valleys, respectively. 
\eqref{axx} and \eqref{axy}
show clearly
the mapping between the noncommutativity produced by the coordinates and 
deformation induced in graphene under the strain effect. Impotently, the mapping  presents an alternative way to give prominence to the experimental realization of the noncommutativity.

On the other hand, 
according to reference \cite{Guinea}
it is convenient to
study the stress tensor 
for the elasticity problems in two-dimensions. Then for a given elastic
energy $\mathcal{F}$ such tensor can be defined as 
$\si_{ij} = \pa \mathcal{F}  / \pa u_{ij}$. Since we have
defined an effective gauge associate to the noncommutative parameter $\kappa$, then immediately one can set
the relations
%
\begin{eqnarray}
 &&
  \frac{\kappa}{2\hbar}y=\pm
  \frac{c\beta}{2a\mu} (\si_{xx} - \si_{yy}) \lb{axx}\\
&& 
- \frac{\kappa}{2\hbar} x
= \mp 2\frac{c\beta }{a\mu} \si_{xy}  \lb{axy}
 \end{eqnarray}
 and $\mu$ is the Lamé coefficient.
 Under some conditions, it was shown  that the elastic energy
has a cubic dependence on the coordinates,  such as
 \cite{Guinea}
 \beq\lb{fff}
 \mathcal{F}(x,y)= c_1 (x+iy)^3 + c_2 (x-iy) (x+iy)^2
 \eeq
 where $c_1$ and $c_2$ are two arbitrary constants.
Note that \eqref{fff} can be worked out to obtain four possible functions
that result in uniform pseudomagnetic field.
%
 Then among them, we choose  the following
\beq\lb{sixy}
\si(x,y)= \si_0 (3x^2y -y^3)
\eeq
with 
$\si_0$ is a constant  depending on the applied  forces.
For the  valley $K$, we evaluate \eqref{axx} and \eqref{axy} 
to establish a link between $\kappa$ and $\si_0$ 
\beq\lb{kasi}
\kappa= \left(12 \frac{c\beta\hbar}{a\mu}\right)  \si_0.
\eeq
To give an explicit value of $\kappa$ as function of $\si_0$ in the case of graphene, we consider the physical parameters 
$\beta \approx 2$,
$a=0.142$ nm, $t=3$ eV,  $c=1$, $\mu = 9.4$ eV Å$^{-2}$. By replacing in \eqref{kasi} 
we offer a measurement of the noncommutative parameter
based on the values taken by $\si_0$. 
Then, in units $\hbar/e = 1$, we end up with
 \beq
\kappa= \left(0.11836412\ 10^{-24}\ \si_0\right) \ \mbox{m s}
\eeq
where $\si_0$ has unit N/m$^2$.
This result may give a hint how to experimentally fix the noncommutative
parameter and therefore offer a measurement of it. Moreover, we can use
other solutions rather than \eqref{sixy} to generate other form of $\kappa$
depending on the geometry of the graphene system.



\section{Conclusion}

We have proposed a method to measure noncommutativity of the phase space variables
using the properties of graphene under the strain effect. More precisely, we have considered Dirac fermions moving in the noncommutative plane and built up a Hamiltonian similar to that describing the same system but experiencing a real
perpendicular magnetic field. This similarity allowed us to define an effective gauge and obtain the effective Landau levels measured in terms of the noncommutative parameter $\kappa$ resulting from the noncommuting momenta.

We have linked the  noncommutativity features to two effects. Indeed, firstly
 we have interpreted $\kappa$ as a manifestation of an effective magnetic field generated by folding space and giving rise to a quantized energy. Secondly, by analogy to pseudovector potential, we have defined $\kappa$ as function of the strain tensor components. To give an explicit form of $\kappa$ in terms of the physical parameters, we have considered the stress tensor and by fixing the physical parameters in the case of graphene, we were be able to give   numerical values of $\kappa$ as function of the stress constant $\si_o$.

\section*{Acknowledgment}
The generous support provided by the Saudi Center for Theoretical
Physics (SCTP) is highly appreciated by AJ.

\end{document}